\newcommand{\zlabel}[1]{\label{#1} }
\newcommand{\roo}{{\rho_1}} %\mbox{\tiny roo:$\roo$}
\newcommand{\fc}{\frac} 
\newcommand{\lt}{\left} 
\newcommand{\rt}{\right} 
\newcommand{\mr}{\mathbf{r}}
\newcommand{\pr}{\prime}

\documentclass[pra,preprint,preprintnumbers,amsmath,amssymb]{revtex4}
\usepackage{graphicx}% Include figure files
\usepackage{dcolumn}% Align table columns on decimal point
\usepackage{epic}%
\usepackage{eepic}%
\usepackage{bm}% bold math
\begin{document}

\preprint{APS/123-QED}

\title{The Differential Virial Theorem with Gradient Formulas for the Operators}

\author{James P. Finley}
\email{james.finley@enmu.edu}
\affiliation{
Department of Physical Sciences,
Eastern New~Mexico University,
Station \#33, Portales, NM 88130}
\date{\today}

\begin{abstract}
A gradient dependent formula is derived for the spinless one-particle density-matrix
operator $\hat{\mathbf{z}}$ from the differential virial theorem. A gradient
dependent formula is also derived for a spinless one-particle density-matrix operator
that can replace the two operators of the differential virial theorem that arise from the
kinetic energy operator. Other operators are also derived that can replace the
operators mentioned above in the differential virial theorem; these operators depend
on the real part of spinless one-particle density-matrix.
\end{abstract}

\maketitle

\section{introduction}

Hohenberg and Kohn \cite{Hohenberg:64} proved that the external potential of an
electronic system is determined by the electron density of the ground
state. Unfortunately, an explicit formula, sequence, or series for this abstract
function with external-potential values has not been discovered. However, an explicit
formula for an operator with external-potential values, but depending on reduced
density matrices of the ground- or excited-states, has been discovered by Holas and
March \cite{Holas:95}. In their formalism, they derived an equation, called
the differential virial theorem, that gives the external potential from a path
integral. The differential virial theorem has been exploited in many theoretical
developments \cite{Qian:98,Sahni:01,Sahni:03,Ryabinkin:12,Nagy:10,Nagy:15,Harbola:98}. The
formalism has also been extended to systems that have a non-integer number of
electrons \cite{Harbola:98} and to ones that are time-dependent \cite{Qian:00}.

For electronic systems, the differential virial theorem is \cite{Holas:95} 
\begin{equation} \zlabel{3820}
\rho(\mathbf{r}) \nabla v(\mathbf{r})
+
2\int 
\rho_2(\mathbf{r},\mathbf{r}^\prime) 
\nabla_{\mathbf{r}}\fc{1}{\mathbf{r}-\mathbf{r}^\pr}
\, d\mathbf{r}^\prime
=
\frac{1}{4} \nabla\nabla^2  \rho(\mathbf{r})
- [\hat{\mathbf{z}}\roo](\mathbf{r}),
\end{equation}
where pair function $\rho_2$, spinless one-particle density matrix $\rho_1$, and
electron density $\rho$ are defined elsewhere \cite{Lowdin:55,McWeeny:60} and below,
and $v$ is a fixed external potential; the value $\hat{\mathbf{z}}\roo$ of the linear
operator $\hat{\mathbf{z}}$ is a vector field with Cartesian components
$\hat{z}_x\roo$, $\hat{z}_y\roo$, and $\hat{z}_z\roo$, where, for example,
$\hat{z}_x\roo$ is defined by
\begin{equation} \zlabel{4825}
[\hat{z}_x\roo](\mathbf{r}) = \fc{1}{2}
\sum_{p\in \{x,y,z\}} \partial_p
\lt[
\lt.\lt(\partial_x\partial_{p^\pr} + \partial_p\partial_{x^\pr}  \rt)
\rho_1(\mathbf{r},\mathbf{r}^\pr)\rt|_{\mathbf{r}^\pr = \mathbf{r}}
\rt],
\end{equation}
and $\partial_{p}$ denotes the partial derivative with respect to
the Cartesian component $p$ of $\mathbf{r}$.
%a similar notation is used for other partial derivatives.

The $\hat{\mathbf{z}}$ term from (\ref{3820}) differs from the others,
since the other terms define a linear operator with a formulas that
depend explicitly on the gradient operator $\nabla$, e.g., let
$\hat{L}$ be defined by
\begin{equation} \zlabel{4728}
\hat{L}\rho(\mr) = \frac{1}{4} \nabla\nabla^2  \rho(\mathbf{r}).
\end{equation}
In this paper, we obtain formulas for the operators $\hat{\mathbf{z}}$ that
depend explicitly on the gradient operator $\nabla$ only. We also obtain other
results as explained within the next two paragraph.

%We also derive other operators that can replace $\hat{\mathbf{z}}$ in (\ref{3820}). We
%also obtain $\nabla$ dependent formulas to replace the right-hand side of (\ref{3820}),
%as explained in the next two paragraph.

Sec.~(\ref{3928}) derives the differential virial theorem. Sec.~(\ref{3930})
obtains equalities that are used in later sections.  Sec.~(\ref{3932}) derives
two linear operators with formulas that depend explicitly on $\nabla$ only, such
that the value $\hat{\mathbf{z}}\roo$ of $\hat{\mathbf{z}}$ at $\roo$ is equal
the value of the derived operators at $\text{Re}\roo$, where $\text{Re}\roo$ is
the real part of $\roo$; so, the derived operators can replace
$\hat{\mathbf{z}}\roo$ in (\ref{3820}), if $\roo$ is replaced by
$\text{Re}\roo$. Sec.~(\ref{3938}) obtains two formulas for $\hat{\mathbf{z}}$
that depend explicitly on $\nabla$ only.

Similar results in Secs.~(\ref{3934}) and (\ref{3938}) are obtained by
treating the operator sum (or difference) defined by right-hand side
of (\ref{3820}). Denoting one of the two derived operators in
Sec.~(\ref{3934}) by $\hat{\kappa}$, we obtain $\hat{L}\rho -
\hat{\mathbf{z}}\roo = \hat{\kappa}\text{Re}\roo$, where $\hat{L}$ is
defined by (\ref{4728}), and $\hat{\kappa}$ has a very simple
formula. An operator, say $\hat{\kappa}_s$, is obtained in
Sec.~(\ref{3938}), such that $\hat{L}\rho - \hat{\mathbf{z}}\roo =
\hat{\kappa}_s\roo$. The formulas for the operators $\hat{\kappa}$ and
$\hat{\kappa}_s$ also depend explicitly on $\nabla$ only.

\section{The Differential Virial Theorem \zlabel{3928}}

Let $\Psi$ be a complex valued, $N$-electron eigenfunction of the time-independent, electronic Schr\"odinger equation
\begin{equation} \zlabel{7372}
\hat{H}\Psi = E\Psi
\end{equation}
where the Hamiltonian operator,
\begin{equation} \zlabel{8272}
\hat{H}  = -\fc12\sum_{i=1}^N \nabla_i^2 + \sum_{i=1}^N v(\mr_i) + \fc12 \sum_{i\ne j}^N r_{ij}^{-1},
\quad r_{ij}^{-1} = |\mathbf{r}_i - \mathbf{r}_j|^{-1},\quad \mathbf{r}\in \mathbb{R}^3
\end{equation}
is determined by the external potential $v$ and the positive integer
$N$. After substituting $\Psi=\chi + i\lambda$ into (\ref{7372}), where
$\chi$ and $\lambda$ are the real and imaginary parts, respectively,
the resulting equation can be separated into two equations
\begin{equation} \zlabel{5820}
\hat{H}\chi = E\chi, \qquad \hat{H}\lambda = E\lambda
\end{equation}
and these are satisfied, since (\ref{7372}) is required to be satisfied by
$\Psi$. Using (\ref{8272}) the equation for the real part is
\begin{equation}
\lt(\sum_{i=1}^N v(\mr_i) + \fc12 \sum_{i\ne j}^N r_{ij}^{-1}\rt)\chi = \fc12\sum_{i=1}^N \nabla_i^2\chi + E\chi.
\end{equation}
%In Cartesian coordinates, $\mathbf{r}_i$ is represented by the ordered triplet $(x_i,y_i,z_i)$. 
Let $\Psi$ have partial derivatives up to the third order that are
continuous for all $\mr$ such that $v(\mr)$ is finite.  Applying the
operator $\chi^2\partial_{x_1}\chi^{-1}$ to the above equation yields
\begin{equation} \zlabel{5924}
\chi^2\partial_{x_1}v(\mr_1) + \chi^2\sum_{j\ne 1}^N \partial_{x_1}r_{1j}^{-1} = 
\fc12\chi^2\sum_{i=1}^N \partial_{x_1}\lt(\chi^{-1}\nabla_i^2\chi\rt).
%\quad \chi(\mr) \text{is finite}
\end{equation}
Applying $\partial_{x_1}$, and then expressing $\nabla_i^2$ in Cartesian
coordinates, the right hand side becomes
\begin{equation} \label{3282} %\zlabel{3282}
\begin{split}
\fc12\chi^2\sum_{i=1}^N \partial_{x}\lt(\chi^{-1}\nabla_i^2\chi\rt)  =&
\fc12\sum_{i=1}^N 
\lt[
\chi\partial_{x_1}\nabla_i^2\chi - (\partial_{x_1}\chi) \nabla_i^2\chi
\rt] 
\\
=&
\sum_{i=1}^N \sum_{p_i\in \{x_i,y_i,z_i\}}
\lt(
\fc12\chi\partial_{p_i}^2\partial_{x_1}\chi - \fc12(\partial_{p_i}^2\chi) \partial_{x_1}\chi 
\rt).
\end{split}
\end{equation}
The identity
\begin{equation}
\fc12 \chi \partial_{p_i}^2 \partial_{x_1} \chi
- \fc12 (\partial_{p_i}^2\chi) \partial_{x_1}\chi
=
\fc14 \partial_{p_i}^2 \partial_{x_1} \chi^2 
- (\partial_{p_i}^2\chi) \partial_{x_1}\chi 
- (\partial_{p_i}\chi)\partial_{p_i}\partial_{x_1}\chi
\end{equation}
is easily proved by, in part, expanding out $\partial_{p_i}^2 \partial_{x_1}
\chi^2$. Using this identity, after substituting (\ref{3282}) into (\ref{5924}),
we find that 
\begin{equation} \zlabel{6702}
\chi^2\partial_{x_1}v(\mr_1) + \chi^2\sum_{j\ne 1}^N \partial_{x_1}r_{1j}^{-1} =
\sum_{i=1}^N \sum_{p_i}%\sum_{p_i\in \{x_i,y_i,z_i\}}
\lt(
\fc14 \partial_{p_i}^2 \partial_{x_1} \chi^2 
- 
\partial_{p_i}\lt[
(\partial_{p_i}\chi) \partial_{x_1}\chi 
\rt]
\rt).
\end{equation}
Eqs.~(\ref{5820}) indicates that (\ref{6702}) with $\chi$ replaced by $\lambda$
is also a true statement. Adding (\ref{6702}) to the corresponding one for
$\lambda$, and using ($|\Psi|^2 = \chi^2 + \lambda^2$), we have
\begin{equation} \zlabel{7472}
|\Psi|^2\partial_{x_1}v(\mr_1) + |\Psi|^2\sum_{j\ne 1}^N \partial_{x_1}r_{1j}^{-1} =
\sum_{i=1}^N \sum_{p_i}%\sum_{p_i\in \{x_i,y_i,z_i\}}
\lt(
\fc14 \partial_{p_i}^2 \partial_{x_1} |\Psi|^2 
- 
\partial_{p_i}\lt[
(\partial_{p_i}\chi) \partial_{x_1}\chi +(\partial_{p_i}\lambda) \partial_{x_1}\lambda 
\rt]
\rt).
\end{equation}
The identity
\begin{equation}
(\partial_{x_1}\chi) \partial_{p_i}\chi  + (\partial_{x_1}\lambda) \partial_{p_i}\lambda =
\fc{1}{2} \lt[ (\partial_{x_1}\Psi) \partial_{p_i}\Psi^*  + (\partial_{p_i}\Psi) \partial_{x_1}\Psi^* \rt]
\end{equation}
is easily proved by substituting $\Psi = \chi + i\lambda$. Hence, (\ref{7472}) can be written
\begin{gather} \label{4820} %\zlabel{4820}
|\Psi|^2\partial_{x_1}v(\mr_1) + |\Psi|^2\sum_{j\ne 1}^N \partial_{x_1}r_{1j}^{-1} 
\hspace{50ex} \\ \notag\hspace{10ex}
=\hspace{0.5ex}\sum_{i=1}^N \sum_{p_i\in \{x_i,y_i,z_i\}} 
\lt\{
\fc14 \partial_{p_i}^2 \partial_{x_1} |\Psi|^2  
-\fc12 
\lt(
\partial_{p_i}
\lt[(\partial_{x_1}\Psi) \partial_{p_i}\Psi^*  + (\partial_{p_i}\Psi) \partial_{x_1}\Psi^*\rt]
\rt) 
\rt\}.
\end{gather}
Equation (\ref{4820}) is integrated over $\mathbf{r}_2, \mathbf{r}_3, \cdots
\mathbf{r}_N$ below. Consider now the integration of the first term on the
right-hand side of (\ref{4820}) over $\mathbf{r}_i$ with the case where
$p_i=x_i$:
\begin{equation} \notag
\begin{split}
\int d\mathbf{r}_i\, \partial_{x_i}^2 \partial_{x_1} (\Psi\Psi^*)  
=
\int\int\int \partial_{x_i}
[(\partial_{x_i}\partial_{x_1}\Psi)\Psi^* 
+ 2(\partial_{x_i}\Psi)\partial_{x_1}\Psi^* 
+ \Psi \partial_{x_i}\partial_{x_1}\Psi^*]
\, dx_i dy_i dz_i 
\\ \hspace{7ex}= 
\lt. 
\int\int 
[(\partial_{x_i}\partial_{x_1}\Psi)\Psi^* 
+ 2(\partial_{x_i}\Psi)\partial_{x_1}\Psi^* 
+ \Psi \partial_{x_i}\partial_{x_1}\Psi^*]
\rt|_{-\infty}^{+\infty}  \, dy_i dz_i = 0
\end{split}
\end{equation}
%
%\begin{equation} \notag
%\begin{split}
%\int d\mathbf{x}_i\, \partial_{x_i}^2 \partial_{x_1} (\Psi\Psi^*)  
%= \int\int\int \partial_{x_i}^2[(\partial_{x_1}\Psi)\Psi^*
%  +\Psi\partial_{x_1}\Psi^*]  \, dx_i dy_i dz_i 
%\hspace{20ex} \\ \hspace{7ex}= 
%\int\int\int \partial_{x_i}
%[(\partial_{x_i}\partial_{x_1}\Psi)\Psi^* 
%+ 2(\partial_{x_i}\Psi)\partial_{x_1}\Psi^* 
%+ \Psi \partial_{x_i}\partial_{x_1}\Psi^*]
%\, dx_i dy_i dz_i =
%\\ \hspace{7ex}= 
%\lt. 
%\int\int 
%[(\partial_{x_i}\partial_{x_1}\Psi)\Psi^* 
%+ 2(\partial_{x_i}\Psi)\partial_{x_1}\Psi^* 
%+ \Psi \partial_{x_i}\partial_{x_1}\Psi^*]
%\rt|_{-\infty}^{+\infty}  \, dy_i dz_i = 0,
%\end{split}
%\end{equation}
%
and this follows since a wave function and its derivative must vanish at
infinities. Obviously, the cases with $p_i=y_i$ and $p_i=z_i$ also
vanish. These results combined with a similar analysis for last terms of
(\ref{4820}) gives
\begin{equation} \notag
\sum_{i=2}^N \sum_{p_i}%\sum_{p_i\in \{x_i,y_i,z_i\}} 
\int d\mathbf{r}_2,d\mathbf{r}_3,\cdots d\mathbf{r}_N \,
\lt\{
\fc14 \partial_{p_i}^2 \partial_{x_1} |\Psi|^2  
-\fc12 
\lt(
\partial_{p_i}
\lt[(\partial_{x_1}\Psi) \partial_{p_i}\Psi^*  + (\partial_{p_i}\Psi) \partial_{x_1}\Psi^*\rt]
\rt) 
\rt\} = 0.
\end{equation}
Note also that
\begin{equation} \notag
(\partial_{x_1}\Psi) \partial_{p_1}\Psi^*  + (\partial_{p_1}\Psi) \partial_{x_1}\Psi^*
\!=\! 
\lt.
(\partial_{x_1}\partial_{p_1^\pr} +  \partial_{p_1}\partial_{x_1^\pr})
\Psi(\mathbf{r_1},\omega_1,\mathbf{x_2},\cdots \mathbf{x_N}) 
\Psi^*(\mathbf{r_1^\pr},\omega_1,\mathbf{x_2},\cdots \mathbf{x_N})
\rt|_{\mathbf{r}_1^\pr = \mathbf{r}_1}
\end{equation}
where $\mathbf{x_i} = \mathbf{r_i},\omega_i$ and $\omega_i$ is the spin
coordinate of the $i$th electron, and the $p_1$ and $p_1^\pr$ components have an
obvious correspondence, e.g., if $p_1=y_1$ then $p_1^\pr=y_1^\pr$. Also, the
notation $\mathbf{r}_1^\pr = \mathbf{r}_1$ means that $\mathbf{r}_1^\pr$ is set
equal to $\mathbf{r}_1$ after all operators have been applied.

Multiplying (\ref{4820}) by $N$, integrating and summing the resulting
equation over $\mathbf{x}_2,\mathbf{x}_3,\cdots \mathbf{x}_N$ and $\omega_1$,
and using the above two identities, we obtain 
\begin{equation} \zlabel{4830}
\rho(\mathbf{r}_1) \partial_{x_1}v(\mathbf{r}_1)
+
2\int (\partial_{x_1} r_{12}^{-1}) \rho_2(\mathbf{r}_1,\mathbf{r}_2) \, d\mathbf{r}_2 
=
\frac{1}{4}\nabla_1^2 \partial_{x_1} \rho(\mathbf{r}_1)
- [\hat{z}_x\roo](\mathbf{r}_1)
\end{equation}
where the value $\hat{z}_x\roo$ of the linear operator $\hat{z}_x$ is defined by (\ref{4825}),
%\begin{equation} \zlabel{4825}
%[\hat{z}_x\roo](\mathbf{r}) = \fc{1}{2}
%\sum_{p\in \{x,y,z\}} \partial_p
%\lt[
%\lt.\lt(\partial_x\partial_{p^\pr} + \partial_p\partial_{x^\pr}  \rt)
%\rho_1(\mathbf{r},\mathbf{r}^\pr)\rt|_{\mathbf{r}^\pr = \mathbf{r}}
%\rt]
%\end{equation}
and it is understood that the summation over $p$ is actually a sum
over the set of ordered pairs $\{(x,x^\pr), (y,y^\pr),(z,z^\pr)\}$; the
spinless one-particle density matrix $\roo$, pair function
$\rho_2$, and density $\rho$ are defined by
\begin{align*}
\rho_1(\mathbf{r}_1,\mathbf{r}_1^\pr) =&\,
N \sum_{\omega_1\cdots\omega_N} \int \Psi(\mr_1,\omega_1,\mathbf{x}_2,\mathbf{x}_3,\cdots \mathbf{x}_N) 
\Psi^*(\mr_1^\pr,\omega_1,\mathbf{x}_2,\mathbf{x}_3,\cdots \mathbf{x}_N) 
\, d\mathbf{r}_2 d\mathbf{r}_3 \cdots d\mathbf{r}_N,
\\
\rho_2(\mathbf{r}_1,\mathbf{r}_2) =&\,
\fc{N(N-1)}{2} \sum_{\omega_1\cdots\omega_N} \int |\Psi|^2
\, d\mathbf{r}_3 d\mathbf{r}_4\cdots d\mathbf{r}_N, \qquad \rho(\mr) = \rho_1(\mr,\mr).
\end{align*}
Obviously, analogous equations to (\ref{4825}) and (\ref{4830})  are
satisfied by the other two components of $\mr_1$, e.g.,, one equation is obtained by replacing $x$
and $x_1$ by $y$ and $y_1$. All three equations combined, after a
change in notation, is the differential virial theorem (\ref{3820}).
%\begin{equation} \zlabel{3820}
%\rho(\mathbf{r}) \nabla v(\mathbf{r})
%+
%2\int 
%\rho_2(\mathbf{r},\mathbf{r}^\prime) 
%\nabla_{\mathbf{r}}\fc{1}{\mathbf{r}-\mathbf{r}^\pr}
%\, d\mathbf{r}^\prime
%=
%\frac{1}{4} \nabla\nabla^2  \rho(\mathbf{r})
%- [\hat{\mathbf{z}}\roo](\mathbf{r}),
%\end{equation}
%the differential virial theorem, where the value $\hat{\mathbf{z}}\roo$ of the linear operator 
%$\hat{\mathbf{z}}$ is a vector field with components $\hat{z}_x\roo$, $\hat{z}_y\roo$, and $\hat{z}_z\roo$.

\section{Equalities involving symmetric operators and functions \zlabel{3930}}

Let $\hat{O}$ be an operator (not necessarily linear) from a
subspace $\mathbb{D}$ of complex valued functions with domain
$D\subset\mathbb{R}^3\times\mathbb{R}^3$. Let $\tilde{\phi}$ denote the real part of a
function $\phi\in\mathbb{D}$, and let $(\mr^\pr,\mr),(\mr,\mr^\pr)\in D$.

If $\hat{O}$ is linear and symmetric, i.e., $\hat{O}_{\mr,\mr^\pr} =
\hat{O}_{\mr^\pr,\mr}$, (e.g., $\hat{O}_{\mr,\mr^\pr} = \nabla_{\mr}\cdot\nabla_{\mr^\pr}$), and
if $\phi^*(\mr,\mr^\pr) = \phi(\mr^\pr,\mr)$, then
\begin{equation} \zlabel{3822}
\lt.\hat{O}_{\mr,\mr^\pr}\phi(\mr,\mr^\pr)\rt|_{\mr^\pr = \mr} 
= \lt.\hat{O}_{\mr,\mr^\pr}\phi^*(\mr,\mr^\pr)\rt|_{\mr^\pr = \mr} 
= \lt.\hat{O}_{\mr,\mr^\pr}\tilde{\phi}(\mr,\mr^\pr)\rt|_{\mr^\pr = \mr}
\end{equation}
where, strictly speaking,
$\hat{O}_{\mr,\mr^\pr}\tilde{\phi}(\mr,\mr^\pr)$ is a short-hand
notation for $\hat{O}_{\mr,\mr^\pr}(\tilde{\phi}(\mr,\mr^\pr)+ i0)$,
and similar notations are used below. The first equality follows from 
\[
\lt.\hat{O}_{\mr,\mr^\pr}\phi(\mr,\mr^\pr)\rt|_{\mr^\pr = \mr} 
= \lt.\hat{O}_{\mr^\pr,\mr}\phi(\mr^\pr,\mr)\rt|_{\mr^\pr = \mr},  
\]
$\hat{O}_{\mr^\pr,\mr} = \hat{O}_{\mr,\mr^\pr}$, and $\phi(\mr^\pr,\mr) =
\phi^*(\mr,\mr^\pr)$. The second one follows from the first one, $\tilde{\phi} =
\fc12(\phi + \phi^*)$, and the requirement that $\hat{O}$ is linear. (Note that
the composite operator defined by the left-hand side of (\ref{3822})---including the $|_{\mr^\pr =
  \mr}$ part---is linear, since $\hat{O}$ is linear.)

If $\phi$ is symmetric, i.e., $\phi(\mr,\mr^\pr)=\phi(\mr^\pr,\mr)$, then
\begin{equation} \zlabel{5294}
\lt.\hat{O}_{\mr,\mr^\pr}\phi(\mr,\mr^\pr)\rt|_{\mr^\pr = \mr} =
\fc12\lt.(\hat{O}_{\mr,\mr^\pr}+\hat{O}_{\mr^\pr,\mr})\phi(\mr,\mr^\pr)\rt|_{\mr^\pr = \mr}.
\end{equation}
This follows by using $\phi(\mr,\mr^\pr)=\phi(\mr^\pr,\mr)$ within
\[
\lt.\hat{O}_{\mr^\pr,\mr}\phi(\mr,\mr^\pr)\rt|_{\mr^\pr = \mr}
= \lt.\hat{O}_{\mr^\pr,\mr}\phi(\mr^\pr,\mr)\rt|_{\mr^\pr = \mr}
= \lt.\hat{O}_{\mr,\mr^\pr}\phi(\mr,\mr^\pr)\rt|_{\mr^\pr = \mr},
\]
and inspection.  

Note that (\ref{5294}) is not an identity if the 
domain $\mathbb{D}$ of $\hat{O}$ includes functions that are not
symmetrical. Similarly, equalities in (\ref{3822}) are not
identities if there exist a $\phi\in\mathbb{D}$ such that
$\phi^*(\mr,\mr^\pr) \ne \phi(\mr^\pr,\mr)$, even if $\tilde{\phi}$
is replaced by $\text{Re}\phi$, where $\text{Re}$ is an operator such
that $\text{Re}\phi= \tilde{\phi}$.

%If $\hat{O}$ is linear and $\hat{O}\phi = \hat{O}\phi^*$, then
%\begin{equation} \zlabel{4294}
%\hat{O}\phi = \hat{O}\tilde{\phi} \quad \text{and}\quad  
%\hat{O}\, \mbox{Im}\phi = 0,
%\end{equation}
%where $\mbox{Im}\phi$ is the imaginary part of $\phi$.  The first one follows from
%$\hat{O}_{\mr^\pr,\mr}\phi = \fc12\hat{O}_{\mr^\pr,\mr}\phi +
%\fc12\hat{O}_{\mr^\pr,\mr}\phi^*$---which is true because $\hat{O}\phi =
%\hat{O}\tilde{\phi}$ and $\hat{O}$ is linear---and using $\tilde{\phi} = \fc12(\phi+\phi^*)$.

\section{Gradient $\nabla$ formulas for operators}

\subsection{Gradient $\nabla$ formulas for operators that can replace $\hat{\mathbf{z}}$ in Eq.~(\ref{3820})\zlabel{3932}}

In this section we derive an operator that can replace
$\hat{\mathbf{z}}$ in the differential viral theorem
(\ref{3820}). This operator depends explicitly on $\text{Re} \roo$
only, and it has a formulas that depend explicitly on the gradient
operator $\nabla$ only. Another $\nabla$ dependent formula for this
operator is obtained by factoring.

Let $C^3(\Omega)$ be the set of all complex valued functions that are
three times continuously differentiable with domain $\Omega= \{
(\mr,\mr^\pr)\in\mathbb{R}^3\times\mathbb{R}^3:v(\mr),v(\mr^\pr)\in
\mathbb{R} \} $. Let $\roo\in C^3(\Omega)$ satisfy
$\rho_1^*(\mr,\mr^\pr)=\rho_1(\mr^\pr,\mr)$ for all $\mr^\pr$ and
$\mr$, such that $v(\mr),v(\mr^\pr)\in \mathbb{R}$. The set
$C^3(\Omega)$ is a subspace of the $\mathbf{L}^2$ Hilbert space with
the same domain $\Omega$.

Since the operator defined by $\partial_x\partial_{p^\pr} +
\partial_p\partial_{x^\pr}$ is symmetric and $\rho_1^*(\mr,\mr^\pr)=\rho_1(\mr^\pr,\mr)$,
the equalities from (\ref{3822}) are applicable, in particular
\begin{equation} \zlabel{4640}
\lt.\lt(\partial_x\partial_{p^\pr} + \partial_p\partial_{x^\pr}  \rt)
\rho_1(\mathbf{r},\mathbf{r}^\pr)\rt|_{\mathbf{r}^\pr = \mathbf{r}} 
=
\lt.\lt(\partial_x\partial_{p^\pr} + \partial_p\partial_{x^\pr}  \rt)
\tilde{\rho}_1(\mathbf{r},\mathbf{r}^\pr)\rt|_{\mathbf{r}^\pr = \mathbf{r}},
\end{equation} 
where $\tilde{\rho}_1$ is the real part of $\roo$.  Since
$\tilde{\rho}_1(\mathbf{r},\mathbf{r}^\pr) =
\tilde{\rho}_1(\mathbf{r}^\pr,\mathbf{r})$, $\tilde{\rho}_1$ is
symmetrical, so we can apply (\ref{5294}) to the right-hand side of
(\ref{4640}), after it is multiplied by $1/2$, giving
\begin{equation} \zlabel{3628}
\fc{1}{2}\lt.\lt(\partial_x\partial_{p^\pr} + \partial_p\partial_{x^\pr}  \rt)
\rho_1(\mathbf{r},\mathbf{r}^\pr)\rt|_{\mathbf{r}^\pr = \mathbf{r}}  =
\lt. \partial_{x}\partial_{p^\pr}\tilde{\rho}_1(\mathbf{r},\mathbf{r}^\pr) \rt|_{\mathbf{r}^\pr = \mathbf{r}}.
\end{equation}
Operating with $\partial_{p}$, we get
\begin{gather} \notag
\fc{1}{2}
\partial_{p}
\lt[
\lt.\lt(\partial_x\partial_{p^\pr} + \partial_p\partial_{x^\pr}  \rt)
\rho_1(\mathbf{r},\mathbf{r}^\pr)\rt|_{\mathbf{r}^\pr = \mathbf{r}}\rt]  =
\partial_{p}
\lt[
\lt. \partial_{x}\partial_{p^\pr}\tilde{\rho}_1(\mathbf{r},\mathbf{r}^\pr) \rt|_{\mathbf{r}^\pr = \mathbf{r}}
\rt]
=
\lt. 
(\partial_{p} + \partial_{p^\pr})
\partial_{x}\partial_{p^\pr}\tilde{\rho}_1(\mathbf{r},\mathbf{r}^\pr) \rt|_{\mathbf{r}^\pr = \mathbf{r}}
\\ =
\lt. 
\lt(
\partial_{p}\partial_{x}\partial_{p^\pr} +
\partial_{p^\pr}\partial_{x}\partial_{p^\pr}
\rt)
\tilde{\rho}_1(\mathbf{r},\mathbf{r}^\pr) 
\rt|_{\mathbf{r}^\pr = \mathbf{r}}
=
\lt. 
\partial_{x}
\lt(
\partial_{p}\partial_{p^\pr} +
\partial_{p^\pr}\partial_{p^\pr}
\rt)
\tilde{\rho}_1(\mathbf{r},\mathbf{r}^\pr) 
\rt|_{\mathbf{r}^\pr = \mathbf{r}}.
\end{gather}
%Summing over $p$, using (\ref{4620}), and comparing with (\ref{4825}), 
Summing over $p$ and comparing with (\ref{4825}), 
we find that 
\begin{equation}
[\hat{z}_x\rho_1](\mathbf{r}) =
\lt. 
\partial_{x}
\lt(
\nabla_{\mathbf{r}}\cdot\nabla_{\mathbf{r}^\pr} + \nabla_{\mathbf{r}^\pr}^2
\rt)
\tilde{\rho}_1(\mathbf{r},\mathbf{r}^\pr) 
\rt|_{\mathbf{r}^\pr = \mathbf{r}}, 
%\quad \tilde{\rho}_1 = (\rho_1+\rho_1^*)/2.
\end{equation}
and by considering all components, we obtain our main objective:
\begin{equation} \label{5730}
[\hat{\mathbf{z}}\rho_1](\mathbf{r}) =
\lt. 
\nabla_\mr
\lt(
\nabla_{\mathbf{r}}\cdot\nabla_{\mathbf{r}^\pr} + \nabla_{\mathbf{r}^\pr}^2
\rt)
\tilde{\rho}_1(\mathbf{r},\mathbf{r}^\pr) 
\rt|_{\mathbf{r}^\pr = \mathbf{r}}, 
\quad \roo\in C^3(\Omega),\quad \rho_1(\mr^\pr,\mr) = \rho_1^*(\mr,\mr^\pr),
\end{equation}
and note that the gradient $\nabla_{\mathbf{r}}$ on the far left is
applied before $\mathbf{r}_1^\pr$ is set equal to $\mathbf{r}_1$.

Let the operator $\hat{\mathbf{z}}^\pr$ be defined by the right-hand
side of the above equation for \emph{all} $\tilde{\rho}_1\in
C^3(\Omega)$ such that $\tilde{\rho}_1$ is real valued. The operator
$\hat{\mathbf{z}}^\pr$ can replace $\hat{\mathbf{z}}$ in (\ref{3820}),
since all one-particle density matrices satisfy $\rho_1(\mr^\pr,\mr) =
\rho_1^*(\mr,\mr^\pr)$.

(Let the operator $\hat{\mathbf{z}}^{\pr\pr}$ be defined by the
right-hand side of the above equation with $\tilde{\rho}_1$ replaced
by $\text{Re}\roo$, where $\text{Re}$ is considered an operator that
is part of the formula for $\hat{\mathbf{z}}^{\pr\pr}$, i.e.,
$\hat{\mathbf{z}}^{\pr\pr}$ is a composition of operators that include
$\text{Re}$.  Note that $\hat{\mathbf{z}}^{\pr\pr} \ne \hat{\mathbf{z}}$ on
$C^3(\Omega)$. However, (\ref{5730}) becomes an identity if the two
operators are restricted to the subspace of functions $\roo$ defined
by the requirements to the right of the same equation.)

Another very similar formula to (\ref{5730}) is obtained by substituting the result from
\begin{gather}
\nabla_{\mathbf{r}}
\lt(
\nabla_{\mathbf{r}}\cdot\nabla_{\mathbf{r}^\pr} + \nabla_{\mathbf{r}^\pr}^2
\rt) 
=
\nabla_{\mathbf{r}}
\lt(
\nabla_{\mathbf{r}^\pr}\cdot\nabla_{\mathbf{r}} + \nabla_{\mathbf{r}^\pr}\cdot\nabla_{\mathbf{r}^\pr}
\rt)
=
\nabla_{\mathbf{r}} \nabla_{\mathbf{r}^\pr}\cdot
\lt(
\nabla_{\mathbf{r}} + \nabla_{\mathbf{r}^\pr}
\rt),
\end{gather}
into (\ref{5730}), giving 
\begin{equation} \zlabel{4720}
%\lt. 
%\nabla_{\mathbf{r}}
%\lt(
%\nabla_{\mathbf{r}}\cdot\nabla_{\mathbf{r}^\pr} + \nabla_{\mathbf{r}^\pr}^2
%\rt)
%\tilde{\rho}_1(\mathbf{r},\mathbf{r}^\pr) 
%\rt|_{\mathbf{r}^\pr = \mathbf{r}}
[\hat{\mathbf{z}}\rho_1](\mathbf{r}) =
\lt. 
\nabla_{\mathbf{r}} \nabla_{\mathbf{r}^\pr}\cdot
\lt(
\nabla_{\mathbf{r}} + \nabla_{\mathbf{r}^\pr}
\rt)
\tilde{\rho}_1(\mathbf{r},\mathbf{r}^\pr) 
\rt|_{\mathbf{r}^\pr = \mathbf{r}}.
\end{equation}
For later use we substitute this result into (\ref{3820}) and then expand out the last term:
%\begin{equation} \label{4030} %\zlabel{4030}
%\begin{split}
\begin{gather} \notag
\rho(\mathbf{r}) \nabla v(\mathbf{r})
+
2\int 
\rho_2(\mathbf{r},\mathbf{r}^\prime) 
\nabla_{\mathbf{r}}\fc{1}{\mathbf{r}-\mathbf{r}^\pr}
\, d\mathbf{r}^\prime
=
\frac{1}{4} \nabla\nabla^2  \rho(\mathbf{r})
- \lt. 
\nabla_{\mathbf{r}} \nabla_{\mathbf{r}^\pr}\cdot
\lt(
\nabla_{\mathbf{r}} + \nabla_{\mathbf{r}^\pr}
\rt)
\tilde{\rho}_1(\mathbf{r},\mathbf{r}^\pr) 
\rt|_{\mathbf{r}^\pr = \mathbf{r}}.
\\ \label{4030}
= 
\frac{1}{4} \nabla\nabla^2  \rho(\mathbf{r})
- \lt. 
\nabla_{\mathbf{r}} \nabla_{\mathbf{r}^\pr}\cdot
\nabla_{\mathbf{r}} 
\tilde{\rho}_1(\mathbf{r},\mathbf{r}^\pr) 
\rt|_{\mathbf{r}^\pr = \mathbf{r}}
- \lt. 
\nabla_{\mathbf{r}} 
\nabla_{\mathbf{r}^\pr}^2
\tilde{\rho}_1(\mathbf{r},\mathbf{r}^\pr) 
\rt|_{\mathbf{r}^\pr = \mathbf{r}}.
\end{gather}

Note that (\ref{4640}) and (\ref{4825})---along with the equations for
$\hat{z}_y\rho_1$ and $\hat{z}_z\rho_1$---indicate that
\begin{equation} \zlabel{4620}
\hat{\mathbf{z}}\roo = \hat{\mathbf{z}}\tilde{\rho}_1.
\end{equation} 
Note also that if we operate on (\ref{3628}) with $\sum_p\partial_{p}$, and then compare with
(\ref{4825}), we obtain
\begin{equation} \zlabel{3620}
[\hat{z}_x\rho_1](\mathbf{r}) = 
\sum_{p\in \{x,y,z\}}\partial_{p}
\lt[
\lt. \partial_{x}\partial_{p^\pr}\tilde{\rho}_1(\mathbf{r},\mathbf{r}^\pr) \rt|_{\mathbf{r}^\pr = \mathbf{r}}
\rt].
\end{equation}

\subsection{Gradient $\nabla$ formulas for operators the can replace the kinetic energy part in Eq.~(\ref{3820})\zlabel{3934}}

Next we show that the top right-hand-side of (\ref{4030})---the
kinetic-energy corresponding part---satisfies
\begin{equation} \label{5524}
\begin{split}
\frac{1}{4} \nabla\nabla^2  \rho(\mathbf{r})
- 
\lt. 
\nabla_{\mathbf{r}} \nabla_{\mathbf{r}^\pr}\cdot
\lt(
\nabla_{\mathbf{r}} + \nabla_{\mathbf{r}^\pr}
\rt)
\tilde{\rho}_1(\mathbf{r},\mathbf{r}^\pr) 
\rt|_{\mathbf{r}^\pr = \mathbf{r}}
 \hspace{30ex}\\ =
\lt.\nabla_{\mr}
\lt(
\mbox{$\frac{1}{2}$}\nabla^2_{\mr} - \mbox{$\frac{1}{2}$}\nabla_{\mathbf{r}^\pr}^2
\rt)
\tilde{\rho}_1(\mr,\mr^\pr)\rt|_{\mr^\pr=\mr} 
 = 
\lt.
\lt(\nabla_{\mr} - \nabla_{\mr^\pr}\rt)
\mbox{$\frac{1}{2}$}\nabla^2_{\mr}\tilde{\rho}_1(\mr,\mr^\pr) 
\rt|_{\mr^\pr=\mr}.
%\hspace{5ex}
\end{split} 
\end{equation}

First we derive an expression for the $\rho$ dependent term in
(\ref{5524}).  Using a set of natural orbitals \cite{Lowdin:55,Parr:89}
$\{\chi_i\}$, the electron density can be written $\rho=
n_i\chi_i\chi_i^*$, where, to reduce clutter, triple indices are
summed over the positive integers. Using the natural orbital
expansion, we have
\begin{equation}
\nabla^2\rho = 
  n_i\nabla\cdot[\chi_i^* \nabla\chi_i]
+ n_i\nabla\cdot[\chi_i\nabla\chi_i^*]
\end{equation}
Using the identity \cite{Arfken:85}
\begin{equation*} 
\nabla\cdot(f\mathbf{V}) = (\nabla f)\cdot\mathbf{V} + f\nabla\cdot\mathbf{V}   
\end{equation*}
where $f$ and $\mathbf{V}$ are scalar and vector functions,
respectively, we obtain
\begin{equation*}
\nabla^2\rho = 
n_i\nabla \chi_i^*\cdot \nabla\chi_i + n_i\chi_i^*\nabla^2\chi_i + \text{cc}
\end{equation*}
where cc means the complex conjugate of the terms to the left.
Operating with $\nabla$ gives
\begin{equation*}
\begin{split}
\nabla\nabla^2\rho &= 
n_i\nabla[\nabla \chi_i^*\cdot \nabla\chi_i] + n_i\nabla[\chi_i^*\nabla^2\chi_i] + \text{cc}
\\ &=
n_i\nabla[\nabla \chi_i^*\cdot \nabla\chi_i]
+ n_i(\nabla\chi_i^*)\nabla^2\chi_i 
+ n_i\chi_i^*\nabla\nabla^2\chi_i 
+ \text{cc}.
\end{split}
\end{equation*}
The value of $\nabla\nabla^2\rho$ at $\mr$ can be expressed by
\begin{equation*}
\nabla_{\mr}\nabla_{\mr}^2\rho(\mr) = 
n_i\nabla_{\mr}[\nabla_{\mr} \chi_i^*(\mr)\cdot \nabla_{\mr}\chi_i(\mr)]
+ \lt.n_i\nabla_{\mr}\nabla_{\mr^\pr}^2 \chi_i(\mr^\pr) \chi_i^*(\mr)\rt|_{\mr^\pr=\mr}   
+ \lt.n_i\nabla_{\mr}\nabla^2_{\mr} \chi_i(\mr)  \chi_i^*(\mr^\pr)\rt|_{\mr^\pr=\mr} 
+ \text{cc}. 
\end{equation*}
Since the first term and the right-hand side is equal to its complex conjugate,
\[
n_i\chi_i(\mr^\pr) \chi_i^*(\mr) + \text{c}c
= \roo(\mr^\pr,\mr) + \text{c}c
=  2 \tilde{\rho}_1(\mr^\pr,\mr),
\]
and $\tilde{\rho}_1(\mr^\pr,\mr) = \tilde{\rho}_1(\mr,\mr^\pr)$, we can write
\begin{equation} \zlabel{2024}
\nabla_{\mr}\nabla_{\mr}^2\rho(\mr) 
= 2n_i\nabla_{\mr}[\nabla_{\mr} \chi_i^*(\mr)\cdot \nabla_{\mr}\chi_i(\mr)]
+ 2\lt.\nabla_{\mr}\nabla_{\mr^\pr}^2 \tilde{\rho}_1(\mr,\mr^\pr)\rt|_{\mr^\pr=\mr}   
+ 2\lt.\nabla_{\mr}\nabla^2_{\mr} \tilde{\rho}_1(\mr,\mr^\pr)\rt|_{\mr^\pr=\mr}. 
\end{equation}

Next we work on first term on the right-hand side of the above equation, starting with
\[
\nabla_{\mr} \chi_i^*(\mr)\cdot \nabla_{\mr}\chi_i(\mr) =
\lt.\nabla_{\mr}\cdot\nabla_{\mr^{\prime}} \chi_i(\mr)\chi_i^*(\mr^{\prime})\rt|_{\mr^\pr=\mr}.
\]
Operating with $\nabla_{\mr}$ gives
\begin{equation} \zlabel{2392}
\nabla_{\mr}[\nabla_{\mr} \chi_i^*(\mr)\cdot \nabla_{\mr}\chi_i(\mr)] =
\lt.\nabla_{\mr}\nabla_{\mr}\cdot\nabla_{\mr^{\prime}} \chi_i(\mr)\chi_i^*(\mr^{\prime})\rt|_{\mr^\pr=\mr} 
+
\lt.\nabla_{\mr^\pr}\nabla_{\mr}\cdot\nabla_{\mr^{\prime}} \chi_i(\mr)\chi_i^*(\mr^{\prime})\rt|_{\mr^\pr=\mr}. 
\end{equation}
%{\bf Check}:
%\begin{gather*}
%\lt.\nabla_{\mr}\cdot\nabla_{\mr^{\prime}} \chi_i(\mr)\chi_i^*(\mr^{\prime})\rt|_{\mr^\pr=\mr} 
%= \sum_p [\partial_{p}\chi_i(\mr)]\partial_{p}\chi_i^*(\mr) 
%\\
%\lt\{\nabla_{\mr} 
%\lt[
%\lt.\nabla_{\mr}\cdot\nabla_{\mr^{\prime}} \chi_i(\mr)\chi_i^*(\mr^{\prime})\rt|_{\mr^\pr=\mr} 
%\rt]\rt\}_x
%= \sum_p [\partial_x\partial_{p}\chi_i(\mr)]\partial_{p}\chi_i^*(\mr) 
%+ \sum_p [\partial_{p}\chi_i(\mr)]\partial_x\partial_{p}\chi_i^*(\mr) 
%\\ \\
%\nabla_{\mr}\cdot\nabla_{\mr^{\prime}} \chi_i(\mr)\chi_i^*(\mr^{\prime})
%= \sum_p\partial_{p}\chi_i(\mr)\partial_{p^\pr}\chi_i^*(\mr^{\prime})
%\\
%\lt\{
%\lt.\nabla_{\mr}\nabla_{\mr}\cdot\nabla_{\mr^{\prime}} \chi_i(\mr)\chi_i^*(\mr^{\prime})\rt|_{\mr^\pr=\mr}
%\rt\}_x  =
%\sum_p [\partial_x\partial_{p}\chi_i(\mr)]\partial_{p}\chi_i^*(\mr) 
%\\
%\lt\{
%\lt.\nabla_{\mr^\pr}\nabla_{\mr}\cdot\nabla_{\mr^{\prime}} \chi_i(\mr)\chi_i^*(\mr^{\prime})\rt|_{\mr^\pr=\mr}
%\rt\}_x  =
%\sum_p [\partial_{p}\chi_i(\mr)]\partial_x\partial_{p}\chi_i^*(\mr) 
%\end{gather*}
For the first term on the right-hand side, $\mr$ and $\mr^\pr$ can be interchanged in
$\nabla_{\mr}\cdot\nabla_{\mr^{\prime}}$ without changing the value; for the second
term, each $\mr$ can be changed to $\mr^\pr$ to the left $|_{\mr^\pr=\mr}$, and,
simultaneously, each $\mr^\pr$ is changed to $\mr$:
\begin{align*}
\lt.\nabla_{\mr}\nabla_{\mr}\cdot\nabla_{\mr^{\prime}} \chi_i(\mr)\chi_i^*(\mr^{\prime})\rt|_{\mr^\pr=\mr}  &=
\lt.\nabla_{\mr}\nabla_{\mr^{\prime}}\cdot\nabla_{\mr} \chi_i(\mr)\chi_i^*(\mr^{\prime})\rt|_{\mr^\pr=\mr},
\\
\lt.\nabla_{\mr^\pr}\nabla_{\mr}\cdot\nabla_{\mr^{\prime}} \chi_i(\mr)\chi_i^*(\mr^{\prime})\rt|_{\mr^\pr=\mr} &=
\lt.\nabla_{\mr}\nabla_{\mr^\pr}\cdot\nabla_{\mr} \chi_i(\mr^\pr)\chi_i^*(\mr)\rt|_{\mr^\pr=\mr}.
\end{align*}
Using these identities after multiplying (\ref{2392}) by $n_i$, we obtain
\begin{align*}
n_i\nabla_{\mr}[\nabla_{\mr} \chi_i^*(\mr)\cdot \nabla_{\mr}\chi_i(\mr)] &=
 n_i\lt.\nabla_{\mr}\nabla_{\mr^{\prime}}\cdot\nabla_{\mr} 
\lt[\chi_i(\mr)\chi_i^*(\mr^{\prime}) +\chi_i(\mr^\pr)\chi_i^*(\mr)\rt]
\rt|_{\mr^\pr=\mr} 
\\ &=2\lt.\nabla_{\mr}\nabla_{\mr^{\prime}}\cdot\nabla_{\mr} \tilde{\rho}_1(\mr,\mr^{\prime})\rt|_{\mr^\pr=\mr}.
\end{align*}
Substituting this result into (\ref{2024}), we find, after multiplying by $\fc14$, that
\begin{equation} %\zlabel{8024}
\fc14\nabla_{\mr}\nabla_{\mr}^2\rho(\mr) 
= \lt.\nabla_{\mr}\nabla_{\mr^{\prime}}\cdot\nabla_{\mr} \tilde{\rho}_1(\mr,\mr^{\prime})\rt|_{\mr^\pr=\mr}
+ \fc12\lt.\nabla_{\mr}\nabla_{\mr^\pr}^2 \tilde{\rho}_1(\mr,\mr^\pr)\rt|_{\mr^\pr=\mr}   
+ \fc12\lt.\nabla_{\mr}\nabla^2_{\mr} \tilde{\rho}_1(\mr,\mr^\pr)\rt|_{\mr^\pr=\mr}.
\end{equation}

Next, the second line of (\ref{4030}) is copied followed by
substituting the above equality:
\begin{gather} \notag 
\frac{1}{4} \nabla\nabla^2  \rho(\mathbf{r})
- \lt. 
\nabla_{\mathbf{r}} \nabla_{\mathbf{r}^\pr}\cdot
\nabla_{\mathbf{r}} 
\tilde{\rho}_1(\mathbf{r},\mathbf{r}^\pr) 
\rt|_{\mathbf{r}^\pr = \mathbf{r}}
- \lt. 
\nabla_{\mathbf{r}} 
\nabla_{\mathbf{r}^\pr}^2
\tilde{\rho}_1(\mathbf{r},\mathbf{r}^\pr) 
\rt|_{\mathbf{r}^\pr = \mathbf{r}} 
\hspace{20ex}
\\ \hspace{15ex}\notag
=\lt.\nabla_{\mr}\nabla_{\mr^{\prime}}\cdot\nabla_{\mr} \tilde{\rho}_1(\mr,\mr^{\prime})\rt|_{\mr^\pr=\mr}
+ \fc12\lt.\nabla_{\mr}\nabla_{\mr^\pr}^2 \tilde{\rho}_1(\mr,\mr^\pr)\rt|_{\mr^\pr=\mr}   
+ \fc12\lt.\nabla_{\mr}\nabla^2_{\mr} \tilde{\rho}_1(\mr,\mr^\pr)\rt|_{\mr^\pr=\mr} 
\\ \notag
\mbox{}- \lt. 
\nabla_{\mathbf{r}} \nabla_{\mathbf{r}^\pr}\cdot
\nabla_{\mathbf{r}} 
\tilde{\rho}_1(\mathbf{r},\mathbf{r}^\pr) 
\rt|_{\mathbf{r}^\pr = \mathbf{r}}
- \lt. 
\nabla_{\mathbf{r}} 
\nabla_{\mathbf{r}^\pr}^2
\tilde{\rho}_1(\mathbf{r},\mathbf{r}^\pr) 
\rt|_{\mathbf{r}^\pr = \mathbf{r}}
%\\
%= \fc12\lt.\nabla_{\mr}\nabla_{\mr^\pr}^2 \tilde{\rho}_1(\mr,\mr^\pr)\rt|_{\mr^\pr=\mr}   
%+ \fc12\lt.\nabla_{\mr}\nabla^2_{\mr} \tilde{\rho}_1(\mr,\mr^\pr)\rt|_{\mr^\pr=\mr} 
%- \lt. 
%\nabla_{\mathbf{r}} 
%\nabla_{\mathbf{r}^\pr}^2
%\tilde{\rho}_1(\mathbf{r},\mathbf{r}^\pr) 
%\rt|_{\mathbf{r}^\pr = \mathbf{r}}.
\\ \zlabel{2628}=
\fc12\lt.\nabla_{\mr}\nabla^2_{\mr} \tilde{\rho}_1(\mr,\mr^\pr)\rt|_{\mr^\pr=\mr} 
- \fc12\lt. 
\nabla_{\mathbf{r}} 
\nabla_{\mathbf{r}^\pr}^2
\tilde{\rho}_1(\mathbf{r},\mathbf{r}^\pr) 
\rt|_{\mathbf{r}^\pr = \mathbf{r}}
\end{gather}
and the first identity from (\ref{5524}) follows by inspection.
Hence, (\ref{4030}) can be written
\begin{equation} \zlabel{5438}
\rho(\mathbf{r}) \nabla v(\mathbf{r})
+
2\int 
\rho_2(\mathbf{r},\mathbf{r}^\prime) 
\nabla_{\mathbf{r}}\fc{1}{\mathbf{r}-\mathbf{r}^\pr}
\, d\mathbf{r}^\prime
=
\lt.\nabla_{\mr}
\lt(
\mbox{$\fc12$} \nabla^2_{\mr} -\mbox{$\fc12$}\nabla_{\mathbf{r}^\pr}^2
\rt) 
\tilde{\rho}_1(\mr,\mr^\pr)\rt|_{\mr^\pr=\mr}. 
\end{equation}

For the second identity from (\ref{5524}), consider the result from
(\ref{2628}).  Interchanging $\mr$ and $\mr^\pr$ and then using
$\tilde{\rho}_1(\mathbf{r},\mathbf{r}^\pr)=\tilde{\rho}_1(\mathbf{r}^\pr,\mathbf{r})$,
we obtain for the last term
\begin{equation}
\fc12\lt. 
\nabla_{\mathbf{r}} 
\nabla_{\mathbf{r}^\pr}^2
\tilde{\rho}_1(\mathbf{r},\mathbf{r}^\pr) 
\rt|_{\mathbf{r}^\pr = \mathbf{r}}
=
\fc12\lt. 
\nabla_{\mathbf{r}^\pr} 
\nabla_{\mathbf{r}}^2
\tilde{\rho}_1(\mathbf{r}^\pr,\mathbf{r}) 
\rt|_{\mathbf{r}^\pr = \mathbf{r}}
=
\fc12\lt. 
\nabla_{\mathbf{r}^\pr} 
\nabla_{\mathbf{r}}^2
\tilde{\rho}_1(\mathbf{r},\mathbf{r}^\pr) 
\rt|_{\mathbf{r}^\pr = \mathbf{r}}.
\end{equation}
Hence
\[
\fc12\lt.\nabla_{\mr}\nabla^2_{\mr} \tilde{\rho}_1(\mr,\mr^\pr)\rt|_{\mr^\pr=\mr} 
- \fc12\lt. 
\nabla_{\mathbf{r}} 
\nabla_{\mathbf{r}^\pr}^2
\tilde{\rho}_1(\mathbf{r},\mathbf{r}^\pr) 
\rt|_{\mathbf{r}^\pr = \mathbf{r}}
=
\fc12\lt.\nabla_{\mr}\nabla^2_{\mr} \tilde{\rho}_1(\mr,\mr^\pr)\rt|_{\mr^\pr=\mr} 
-\fc12\lt. 
\nabla_{\mathbf{r}^\pr} 
\nabla_{\mathbf{r}}^2
\tilde{\rho}_1(\mathbf{r},\mathbf{r}^\pr) 
\rt|_{\mathbf{r}^\pr = \mathbf{r}}.
\]
Comparing this identity with (\ref{2628}) gives the second identity
from (\ref{5524}), and (\ref{4030}) can be written
\begin{equation}
\rho(\mathbf{r}) \nabla v(\mathbf{r})
+
2\int 
\rho_2(\mathbf{r},\mathbf{r}^\prime) 
\nabla_{\mathbf{r}}\fc{1}{\mathbf{r}-\mathbf{r}^\pr}
\, d\mathbf{r}^\prime
=
\lt.
\lt(\nabla_{\mr} - \nabla_{\mr^\pr}\rt)
\mbox{$\frac{1}{2}$}\nabla^2_{\mr}\tilde{\rho}_1(\mr,\mr^\pr) 
\rt|_{\mr^\pr=\mr}.
\end{equation}

\section{Gradient $\nabla$ formulas for $\hat{\mathbf{z}}$ and the kinetic energy part \zlabel{3938} }

In this section we symmetrize operators derived in the previous two
sections to obtain gradient-dependent formulas for the operator $\hat{\mathbf{z}}$
and a spinless one-particle density-matrix operator that can replace
the right-hand side of (\ref{3820}).

Consider (\ref{4720}), and let
\begin{equation} \zlabel{8720}
\hat{O}_{\mr,\mr^\pr} =
\nabla_{\mathbf{r}} \nabla_{\mathbf{r}^\pr}\cdot
\lt(
\nabla_{\mathbf{r}} + \nabla_{\mathbf{r}^\pr}
\rt)
\end{equation}
Since $\tilde{\rho}_1$ is symmetric, we can use equality (\ref{5294}) in (\ref{4720}):
\[
[\hat{\mathbf{z}}\rho_1](\mathbf{r}) =
\lt.\hat{O}_{\mr,\mr^\pr}\tilde{\rho}_1(\mr,\mr^\pr)\rt|_{\mr^\pr = \mr} =
\fc12\lt.(\hat{O}_{\mr,\mr^\pr}+\hat{O}_{\mr^\pr,\mr})\tilde{\rho}_1(\mr,\mr^\pr)\rt|_{\mr^\pr = \mr}.
\]
Since $\fc12(\hat{O}_{\mr,\mr^\pr}+\hat{O}_{\mr^\pr,\mr})$ is linear
and symmetric, and $\rho_1^*(\mr,\mr^\pr) = \roo(\mr^\pr,\mr)$, we can
use equalities from (\ref{3822}), in particular
\[
\lt.\fc12(\hat{O}_{\mr,\mr^\pr}+\hat{O}_{\mr^\pr,\mr})\roo(\mr,\mr^\pr)\rt|_{\mr^\pr = \mr} 
= \lt.\fc12(\hat{O}_{\mr,\mr^\pr}+\hat{O}_{\mr^\pr,\mr})\tilde{\rho}_1(\mr,\mr^\pr)\rt|_{\mr^\pr = \mr} .
\]
Substituting this result into the one above it, we have
\begin{equation} \zlabel{4028}
[\hat{\mathbf{z}}\rho_1](\mathbf{r}) = 
\lt.\fc12(\hat{O}_{\mr,\mr^\pr}+\hat{O}_{\mr^\pr,\mr})\roo(\mr,\mr^\pr)\rt|_{\mr^\pr = \mr}
\end{equation}
and, explicitly, we obtain one of our objectives from (\ref{8720}):
\begin{equation} \zlabel{7028}
[\hat{\mathbf{z}}\rho_1](\mathbf{r}) =
\lt. 
\fc12
\lt(
  \nabla_{\mathbf{r}} \nabla_{\mathbf{r}^\pr}
+ \nabla_{\mathbf{r}^\pr} \nabla_{\mathbf{r}}
\rt) \cdot
\lt(
\nabla_{\mathbf{r}} + \nabla_{\mathbf{r}^\pr}
\rt)
\rho_1(\mathbf{r},\mathbf{r}^\pr) 
\rt|_{\mathbf{r}^\pr = \mathbf{r}}.
\end{equation}

Consider (\ref{5730}). Using the same procedure as above with
$\hat{O}_{\mr^\pr,\mr}=\nabla_\mr \lt( \nabla_{\mathbf{r}}\cdot\nabla_{\mathbf{r}^\pr} +
\nabla_{\mathbf{r}^\pr}^2 \rt)$,
we also find that 
%(\ref{4028}) is true, and explicitly, 
%we obtain
\begin{equation}
\zlabel{8028}
[\hat{\mathbf{z}}\rho_1](\mathbf{r}) =
\fc12
\lt[
\lt. 
\lt(\nabla_\mr + \nabla_{\mr^\pr}\rt)
\lt(
\nabla_{\mathbf{r}}\cdot\nabla_{\mathbf{r}^\pr}
\rt)
+
\nabla_\mr\nabla_{\mathbf{r}^\pr}^2 + \nabla_{\mr^\pr}\nabla_{\mathbf{r}}^2
\rt]
\rho_1(\mathbf{r},\mathbf{r}^\pr) 
\rt|_{\mathbf{r}^\pr = \mathbf{r}}
\end{equation}
and the two formulas above, differ only by factorization.

Next consider (\ref{5524}), and let the operator $\hat{\kappa}$ and
$\hat{\kappa}^\pr$ have values that are defined by
\begin{align} \zlabel{0724}
[\hat{\kappa}\tilde{\rho}_1](\mathbf{r}) &=
\lt.\nabla_{\mr}
\lt(
\mbox{$\frac{1}{2}$}\nabla^2_{\mr} - \mbox{$\frac{1}{2}$}\nabla_{\mathbf{r}^\pr}^2
\rt)
\tilde{\rho}_1(\mr,\mr^\pr)\rt|_{\mr^\pr=\mr},
\\ \zlabel{0728}
[\hat{\kappa}^\pr\tilde{\rho}_1](\mathbf{r}) &= 
\lt.
\lt(\nabla_{\mr} - \nabla_{\mr^\pr}\rt)
\mbox{$\frac{1}{2}$}\nabla^2_{\mr}\tilde{\rho}_1(\mr,\mr^\pr) 
\rt|_{\mr^\pr=\mr}
\end{align}
and either one of these can replace the right-hand side of the
differential virial theorem (\ref{3820}). Applying the symmetrization
procedure above to $\hat{\kappa}$, we find that
\begin{equation} \zlabel{0720}
[\hat{\kappa}\tilde{\rho}_1](\mathbf{r}) = 
\fc14 \lt.
\lt(\nabla_{\mr} - \nabla_{\mr^\pr}\rt)
\lt(
\nabla^2_{\mr} - \nabla_{\mathbf{r}^\pr}^2
\rt)
\rho_1(\mr,\mr^\pr)\rt|_{\mr^\pr=\mr}.
\end{equation}
%{\bf Check (ignoring factors)}
%\begin{gather*}
%\nabla_{\mr}(\nabla^2_{\mr} - \nabla_{\mathbf{r}^\pr}^2) +
%\nabla_{\mr^\pr}(\nabla^2_{\mr^\pr} - \nabla_{\mr}^2)
%=
%\nabla_{\mr}(\nabla^2_{\mr} - \nabla_{\mathbf{r}^\pr}^2) -
%\nabla_{\mr^\pr}(\nabla^2_{\mr} - \nabla_{\mr^\pr}^2)
%=
%\lt(\nabla_{\mr} - \nabla_{\mr^\pr}\rt)
%\lt(
%\nabla^2_{\mr} - \nabla_{\mathbf{r}^\pr}^2
%\rt)
%\end{gather*}

Consider the following identities with parts from within (\ref{0724}) and (\ref{0728}): 
\begin{gather*}
\nabla_{\mr}
(\nabla^2_{\mr} - \nabla_{\mathbf{r}^\pr}^2) 
= \nabla_{\mr}\nabla^2_{\mr} - \nabla_{\mr}\nabla_{\mathbf{r}^\pr}^2,
\\
\lt(\nabla_{\mr} - \nabla_{\mr^\pr}\rt)\nabla^2_{\mr}
= \nabla_{\mr}\nabla^2_{\mr} - \nabla_{\mr^\pr}\nabla^2_{\mr}.
\end{gather*}
By comparing these two it is obvious that the symmetrization procedure
involving $\hat{\kappa}^\pr$ gives the same result, i.e.,
$\hat{\kappa}^\pr\tilde{\rho}_1(\mathbf{r})$ is also equal to the
right-hand side of (\ref{0720}). Substituting (\ref{0724}) into
(\ref{5438}) and using (\ref{0720}), we find that 
\begin{equation} %\zlabel{}
\rho(\mathbf{r}) \nabla v(\mathbf{r})
+
2\int 
\rho_2(\mathbf{r},\mathbf{r}^\prime) 
\nabla_{\mathbf{r}}\fc{1}{\mathbf{r}-\mathbf{r}^\pr}
\, d\mathbf{r}^\prime
=
\fc14 \lt.
\lt(\nabla_{\mr} - \nabla_{\mr^\pr}\rt)
\lt(
\nabla^2_{\mr} - \nabla_{\mathbf{r}^\pr}^2
\rt)
\rho_1(\mr,\mr^\pr)\rt|_{\mr^\pr=\mr}.
\end{equation}

%{\bf Check something}
%\begin{gather*}
%\fc12\lt[\nabla_{\mr}
%\lt(
%\mbox{$\frac{1}{2}$}\nabla^2_{\mr} - \mbox{$\frac{1}{2}$}\nabla_{\mathbf{r}^\pr}^2
%\rt)
%+
%\nabla_{\mr^\pr}
%\lt(
%\mbox{$\frac{1}{2}$}\nabla^2_{\mr^\pr} - \mbox{$\frac{1}{2}$}\nabla_{\mathbf{r}}^2
%\rt)\rt]
%= 
%\fc12\lt[\nabla_{\mr}
%\lt(
%\mbox{$\frac{1}{2}$}\nabla^2_{\mr} - \mbox{$\frac{1}{2}$}\nabla_{\mathbf{r}^\pr}^2
%\rt)
%-
%\nabla_{\mr^\pr}
%\lt(
%\mbox{$\frac{1}{2}$}\nabla^2_{\mr} - \mbox{$\frac{1}{2}$}\nabla_{\mathbf{r}^\pr}^2
%\rt)\rt]
%\\
%= \fc12
%\lt(\nabla_{\mr} - \nabla_{\mr^\pr}\rt)
%\lt(
%\mbox{$\frac{1}{2}$}\nabla^2_{\mr} - \mbox{$\frac{1}{2}$}\nabla_{\mathbf{r}^\pr}^2
%\rt)
%\end{gather*}
%%
%\begin{gather*}
%\fc12\lt[
%\lt(\nabla_{\mr} - \nabla_{\mr^\pr}\rt)
%\mbox{$\frac{1}{2}$}\nabla^2_{\mr}
%+
%\lt(\nabla_{\mr^\pr} - \nabla_{\mr}\rt)
%\mbox{$\frac{1}{2}$}\nabla^2_{\mr^\pr}
%\rt]
%= \fc12\lt[
%\lt(\nabla_{\mr} - \nabla_{\mr^\pr}\rt)
%\mbox{$\frac{1}{2}$}\nabla^2_{\mr}
%-
%\lt(\nabla_{\mr} - \nabla_{\mr^\pr}\rt)
%\mbox{$\frac{1}{2}$}\nabla^2_{\mr^\pr}
%\rt]
%\\
%= \fc12\lt[
%\lt(\nabla_{\mr} - \nabla_{\mr^\pr}\rt)
%\lt(
%\mbox{$\frac{1}{2}$}\nabla^2_{\mr}
%- \mbox{$\frac{1}{2}$}\nabla^2_{\mr^\pr}
%\rt)\rt]
%\end{gather*}

\section{Conclusion}

Identity (\ref{7028}) (or (\ref{8028})\hspace{0.3ex}) provides a new
formula for the operator $\hat{\mathbf{z}}$ from the differential
virial theorem (\ref{3820}), previously defined, in part, by
(\ref{4825}). The right-hand side of (\ref{4720}) (or
(\ref{5730})\hspace{0.3ex}) can also replace
$[\hat{\mathbf{z}}\rho_1](\mathbf{r})$ in (\ref{3820}), where
$\tilde{\rho}_1$ is the real part of the spin-less one particle
density matrix $\roo$. The two terms on the right-hand side of
(\ref{3820})---corresponding to the kinetic energy---can be replaced
by the right-hand side of (\ref{0720}); these two terms can also be
replaced by $[\hat{\kappa}\tilde{\rho}_1](\mathbf{r})$, defined by
(\ref{0724}) (or $[\hat{\kappa}^\pr\tilde{\rho}_1](\mathbf{r})$
defined by (\ref{0728})\hspace{0.3ex}).  The new formula for the
operator $\hat{\mathbf{z}}$, the $\roo$-dependent operator for the
kinetic energy part, and the operators that depend on
$\tilde{\rho}_1$, depend on the gradient operator $\nabla$ only.

\bibliography{ref}
\end{document}